\documentclass[aps,prd,reprint,showkeys,nofootinbib,superscriptaddress,notitlepage]{revtex4-1}
\usepackage{amsmath,graphicx,amssymb,xspace}
\usepackage[utf8]{inputenc}
\usepackage[unicode=true,
  bookmarks=false,   backref=false,    colorlinks=true,
  linktocpage=true,  citecolor=black,  linkcolor=black,
  urlcolor=black,    breaklinks=true
]{hyperref}

\newcommand{\lcdm}{$\Lambda$CDM\xspace}
\newcommand{\cpl}{$w_0 w_a$CDM\xspace}
\newcommand{\g}{$g_{\mu\nu}$\xspace}
\newcommand{\f}{$f_{\mu\nu}$\xspace}
\newcommand{\kg}{\kappa_g}
\newcommand{\kf}{\kappa_f}
\newcommand{\mfp}{m_\mathrm{FP}}

\newcommand{\wde}{w_\mathrm{DE}}
\newcommand{\hu}{\, \mathrm{km/s/Mpc}}
\newcommand{\Om}{\Omega_{m,0}}
\newcommand{\Or}{\Omega_{r,0}}
\newcommand{\OmL}{\Omega_\Lambda}
\newcommand{\Ode}{\Omega_\mathrm{DE}}

\newcommand{\Dchi}{\Delta \chi^2_\mathrm{MAP}}

\newcommand{\pp}{Pantheon+\xspace}
\newcommand{\uni}{Union3\xspace}
\newcommand{\des}{DES Y5\xspace}

\begin{document}

\title{Bimetric gravity improves the fit to DESI BAO and eases the Hubble tension}

\author{Marcus \surname{Högås}}
\email{marcus.hogas@fysik.su.se}
\affiliation{Oskar Klein Centre, Department of Physics, Stockholm University\\Albanova University Center\\ 106 91 Stockholm, Sweden}

\author{Edvard \surname{Mörtsell}}
\email{edvard@fysik.su.se}
\affiliation{Oskar Klein Centre, Department of Physics, Stockholm University\\Albanova University Center\\ 106 91 Stockholm, Sweden}

\begin{abstract}
We investigate whether the latest combination of DESI DR2 baryon acoustic oscillation (BAO) measurements, cosmic microwave background (CMB) data (\emph{Planck} 2018 + ACT), and Type~Ia supernovae (SNe~Ia) compilations (Pantheon+, Union3, and DES Y5) favor a dynamical dark energy component, and explore if such a scenario can simultaneously help resolve the Hubble tension. We contrast two frameworks: the widely used phenomenological $w_0 w_a$CDM model, and bimetric gravity---a fundamental modification of general relativity that naturally gives rise to phantom dark energy.
The $w_0 w_a$CDM model is moderately preferred over $\Lambda$CDM, at the $2$--$4 \, \sigma$ level, when fitting DESI DR2 + CMB + SNe~Ia, but it exacerbates the Hubble tension.
By comparison, bimetric gravity provides a modest improvement in fit quality, at the $1 \, \sigma$ level, but, by inferring $H_0 = 69.0 \pm 0.4 \, \mathrm{km/s/Mpc}$, it partially eases the Hubble tension, from a $5 \,\sigma$ discrepancy to a $3.7 \, \sigma$ tension.
Including locally calibrated SNe Ia brings the overall preference for the bimetric model over $\Lambda$CDM to the $2 \, \sigma$ level, comparable to that of the $w_0 w_a$CDM model when including the local SN Ia calibration. 
\end{abstract}

\maketitle

\section{Introduction}
The standard cosmological model, built on general relativity and quantum physics, has a remarkably successful history in predicting and explaining cosmological observations.
The discovery of cosmic acceleration in the late 1990s led to the reintroduction of the cosmological constant ($\Lambda$) as a key component of this framework \cite{1917SPAW.......142E,SupernovaCosmologyProject:1998vns,SupernovaSearchTeam:1998fmf}.
However, new data hints at an evolving dark energy component. In particular, the Dark Energy Spectroscopic Instrument (DESI) Year 3 data release (DR2)---which provides an unprecedented 3D map of more than 14 million galaxies and quasars---has enabled the most precise BAO measurements to date.
These measurements, when combined with complementary data from the CMB and SNe Ia, suggest that the dark energy density may evolve with time \cite{DESI:2025fii,DESI:2025zgx}.

However, the statistical significance of evolving dark energy is still modest---typically $ \lesssim 4 \,\sigma$, depending on the dataset combination and parameterization---and should be interpreted with appropriate caution.
For instance, systematic uncertainties in the SNe Ia can significantly affect the inferred preference for dynamical dark energy \cite{Dhawan:2024gqy,Gialamas:2024lyw,Huang:2025som,Afroz:2025iwo,Dhawan:2025mer}, possibly due to subtle effects in their cross-calibration \cite{Popovic:2025glk}.
On the other hand, a majority of dataset combinations---spanning BAO measurements from DESI and the Sloan Digital Sky Survey (SDSS), CMB data from \emph{Planck}, the Atacama Cosmology Telescope (ACT), and the South Pole Telescope (SPT), and SNe Ia from \pp, \uni, and the Dark Energy Survey Year 5 (\des) data compilations---do favor a time-evolving dark energy. The strength of this preference varies, and a few specific combinations---such as SDSS BAO with \pp SNe Ia, or CMB data from SPT---show no significant evidence for dynamical evolution \cite{Giare:2025pzu,Giare:2025tvz}.

Much of contemporary literature study phenomenological models of dark energy, with the Chevallier--Polarski--Linder (CPL) parameterization being the most widely used, with the equation of state $\wde(a) = w_0 + w_a (1-a)$ as a function of the scale factor, $a$. 
We will refer to this as the \cpl model.
Although computationally efficient, \cpl is not derived from fundamental physics and thus lacks theoretically justified parameter priors. Still, such models can offer guidance on which features valid fundamental theories may need to accommodate.
One feature allowed for in \cpl is phantom crossing, where the dark energy equation of state $\wde (a)$ evolves across the phantom divide $w = -1$ \cite{Ozulker:2025ehg,Gonzalez-Fuentes:2025lei,Scherer:2025esj,RoyChoudhury:2025dhe}.
Realizing such an evolution within a fundamental physics framework such as a scalar field model of dark energy faces theoretical challenges, including instabilities and fine-tuning \cite{Carroll:2003st,Vikman:2004dc}, although there are exceptions \cite{Clifton:2011jh,Hu:2004kh,Ishak:2018his,Koussour:2023ulc,Wolf:2024stt,Gomez-Valent:2025mfl}.

Another possible hint for dynamical dark energy is the Hubble tension---the persistent discrepancy between different estimates of the Hubble constant ($H_0$) which sets the present-day expansion rate of the Universe. CMB observations yield a low value, $H_0 = 67.4 \pm 0.5 \hu$ \cite{Planck:2018vyg}, when interpreted within the \lcdm framework. In contrast, late-Universe measurements, most notably the SH0ES team's distance ladder estimate, give a significantly higher value, $H_0 = 73.0 \pm 1.0 \hu$ \cite{Riess:2021jrx}.
This $8 \, \%$ difference exceeds the $5 \, \sigma$ level and has sparked extensive debate in the literature.
A possible resolution may be unaccounted-for systematics in the distance ladder, see for example \cite{Mortsell:2021nzg,Mortsell:2021tcx,Hogas:2024qlt}.
Another possible solution involves modifying the cosmological expansion history, for example through a dynamical dark energy component that increases the CMB-inferred value of $H_0$. See Refs.~\cite{DiValentino:2021izs,Schoneberg:2021qvd,Abdalla:2022yfr,CosmoVerse:2025txj} for reviews.

Taken together, the DESI DR2 results and the Hubble tension present two potential hints of dynamical dark energy. A natural question is whether they point toward a common form of dark energy. For the CPL parameterization, the answer appears to be no: when fitting DESI DR2 jointly with CMB and SNe Ia data, the inferred Hubble constant falls even below the \lcdm value of $67.4 \hu$, thereby worsening the tension \cite{DESI:2025fii,DESI:2025zgx,Pang:2025lvh}.\\

\noindent \textbf{Executive summary.} In this work, we show that bimetric gravity---a natural extension of general relativity---is preferred over \lcdm at the $1 \, \sigma$ level by DESI DR2 BAO + CMB + SNe~Ia data, while also predicting a higher Hubble constant, $H_0 = 69.0 \pm 0.4 \hu$, thereby easing the Hubble tension to $3.7 \, \sigma$ from an initial $> 5 \, \sigma$ discrepancy. Although current data offer intriguing but not yet decisive hints of dynamical dark energy, bimetric gravity stands out as a theoretically motivated scenario. Future observational efforts, including model-independent reductions of the DESI BAO data and targeted searches for bimetric-specific gravitational-wave signatures, will be essential to determine whether these hints reflect genuine new physics beyond \lcdm.

\section{Data description}
Our analysis incorporates BAO measurements from DESI DR2, derived from observations of galaxies, quasars, and Lyman-$\alpha$ tracers \cite{DESI:2025zgx}.
For the CMB, we use a compressed likelihood combining data from the \emph{Planck} 2018 data release \cite{Planck:2019nip} and ACT Data Release 6 \cite{ACT:2023kun}, as provided in Ref.~\cite{Bansal:2025ipo}.
SNe Ia distances and magnitudes are taken from three sources: \pp \cite{Scolnic:2021amr,Brout:2022vxf}, the binned \uni sample \cite{Rubin:2023jdq}, and \des \cite{DES:2024jxu}. 
To minimize systematic effects due to peculiar velocities, we include only supernovae with $z > 0.023$.\footnote{This differs from the redshift cut employed in the DESI DR2 cosmological analysis where $z > 0.01$.}

\section{Bimetric cosmology}
Bimetric gravity is a natural extension of general relativity (GR) introducing a second metric, denoted \f, alongside the familiar space-time metric of GR, \g.
The requirement of theoretical consistency makes the theory strongly constrained, introducing four constant parameters beyond those of \lcdm \cite{Hassan:2011zd,Hogas:2021fmr}.
When the equations of motion are linearized, the theory reveals two types of gravitational waves: one massless, like in GR, and one massive \cite{Hassan:2011zd,Hassan:2012wr}. The mass is denoted $\mfp$ after Fierz and Pauli \cite{Fierz:1939ix}. 
Importantly, each of the metrics \g and \f is a linear combination of the massless and massive modes. This is similar to how neutrino flavor states are mixtures of mass states. The degree of mixing is controlled by a parameter known as the mixing angle, $\theta$, with GR being recovered as $\theta \to 0$, where the massless mode aligns with the space-time metric \g, whereas for $\theta \to \pi / 2$, it is aligned with the massive mode \cite{Hogas:2021fmr}.

The redshift evolution of the cosmic expansion rate, $H(z)$, is governed by the Friedmann equation. In a spatially flat universe, it reads
\begin{equation}
\label{eq:Friedmann}
    \left( \frac{H(z)}{H_0} \right)^2 = \Om (1+z)^3 + \Or (1+z)^4 + \Ode(z).
\end{equation}
Here, the dark energy density $\Ode(z)$ originates from the interaction between the two metrics and can be expressed in terms of the ratio of the scale factors of the two metrics, $y = a_g / a_f$, with $a_g$ being the scale factor of \g and $a_f$ the scale factor of \f,
\begin{equation}
    \Ode = \OmL - \sin^2 \theta \, \mfp^2 (1-y) \left[ 1 + \alpha (1-y) + \frac{\beta}{3} (1-y)^2 \right].
\end{equation}
The ratio of the scale factors, $y$, is the solution of a quartic polynomial whose coefficients are determined by the theory parameters and the total mass density (see Appendix~\ref{sec:Action_EoM}). Moreover, $y$ is a monotonically increasing function of time: it starts at $y=0$ at the Big Bang and approaches $y=1$ in the far future. Thus, $\OmL$ is the asymptotic (far-future) cosmological-constant value of the dark energy.
In addition to the parameters $\theta$, $\mfp$, and $\OmL$, the theory includes two additional constants, $\alpha$ and $\beta$, described further in Ref.~\cite{Hogas:2021fmr,Hogas:2021lns}, which play a key role in ensuring the presence of a functional Vainshtein screening mechanism restoring GR in high-density environments. 
The bimetric action and cosmological equations of motion are described in detail in Appendix~\ref{sec:Action_EoM} and the numerical procedure for solving the cosmological equations is described in Appendix~\ref{sec:NumSol}. 

\begin{figure}
    \centering
    \includegraphics[width=1\linewidth]{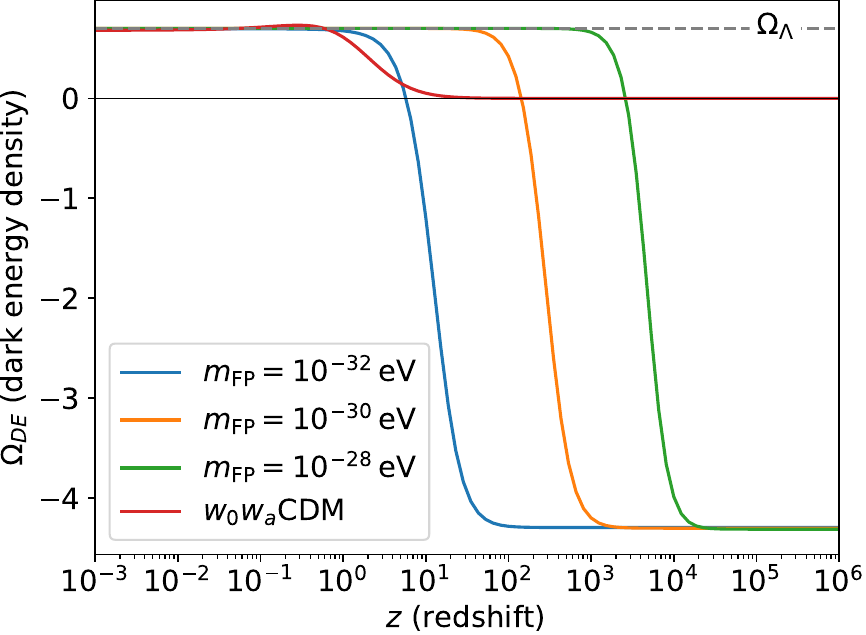}
    \caption{Redshift evolution of the dark energy density for three different values of the bimetric mass parameter $\mfp$. For reference, $H_0 \sim 10^{-33} \, \mathrm{eV}$. While the detailed evolution of $\Ode$ can vary considerably depending on the theory parameters, it always follows the same qualitative pattern: beginning as a (possibly negative) cosmological constant in the early universe, here around $-4$, then transitioning to a greater late-time value $\OmL$. A larger $\mfp$ corresponds to an earlier transition. For comparison, the evolution of a \cpl model with $w_0 = -0.8$ and $w_a = -0.9$ is also shown, highlighting differences in behavior between the bimetric and \cpl models.}
    \label{fig:Omega_DE_Example}
\end{figure}

Bimetric cosmology does not allow for phantom crossing, but it gives rise to a stable phantom dark energy component \cite{Mortsell:2017fog,Hogas:2021fmr,Hogas:2021lns}. 
This makes it a natural framework for testing whether the evolution favored by DESI DR2 and other recent data can be accommodated within a consistent theory.
At early times, the dark energy component typically behaves like a (possibly negative) cosmological constant, then transitions through a dynamical phase, and eventually settles into a larger, positive cosmological constant $\OmL$. The redshift at which this transition occurs is governed by the mass of the gravitational field ($\mfp$), see Fig.~\ref{fig:Omega_DE_Example} for examples.

Altogether, the theory contains five parameters---four more than the cosmological constant already present in GR.
Crucially, to ensure a ghost-free cosmology and a functioning Vainshtein screening mechanism---which is essential for recovering GR on solar-system and galactic scales, and thus for observational viability---we explicitly impose these constraints in our statistical analysis. The detailed conditions are listed in Ref.~\cite{Hogas:2021fmr}.

\section{Results}
Recently, Ref.~\cite{Smirnov:2025yru} analyzed a set of bimetric models, comparing their predicted equation of state with a Gaussian Process reconstruction from Ref.~\cite{DESI:2025fii}.
In the present work, we go further by performing a full statistical data analysis of the most general bimetric model.
The posterior distribution for the model parameters is inferred using the \texttt{emcee} Python library, implementing Goodman and Weare's affine invariant Markov Chain Monte Carlo ensamble sampling algorithm \cite{Goodman:2010dyf,Foreman-Mackey:2012any}. The likelihood is described in detail in Appendix~\ref{sec:NumSol}.

\begin{figure}
    \centering
    \includegraphics[width=0.85\linewidth]{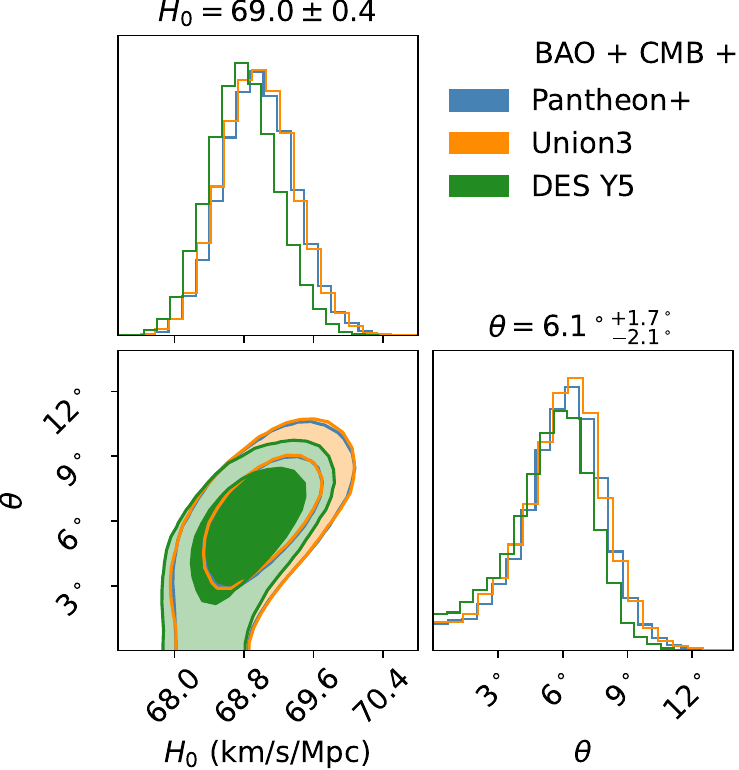}
    \caption{Results for a general bimetric model. Marginalized posterior constraints on $H_0$ and $\theta$ from BAO (DESI DR2), CMB (\emph{Planck} 2018 + ACT), and three different SNe Ia compilations. Contours show the $68 \, \%$ and $95 \, \%$ credence regions. In the limit $\theta \to 0$, the standard \lcdm model is recovered. In contrast to the \cpl model, the results show strong consistency across the three SNe Ia compilations. The inferred Hubble constant is significantly higher than for the \cpl and \lcdm models, thereby reducing the Hubble tension from the $5 \, \sigma$ level to $3.7 \, \sigma$.}
    \label{fig:H0_theta}
\end{figure}

In Fig.~\ref{fig:H0_theta} we show the results for $H_0$ and $\theta$ when a general bimetric model (five free theory parameters) is fitted to BAO (DESI DR2), CMB (\emph{Planck} 2018 + ACT), and three different SNe Ia compilations. Recall that \lcdm is recovered in the $\theta \to 0$ limit. We observe a clear preference for a non-zero $\theta$ indicating that the bimetric model provides a better fit than \lcdm.

\begin{figure}
    \centering
    \includegraphics[width=1\linewidth]{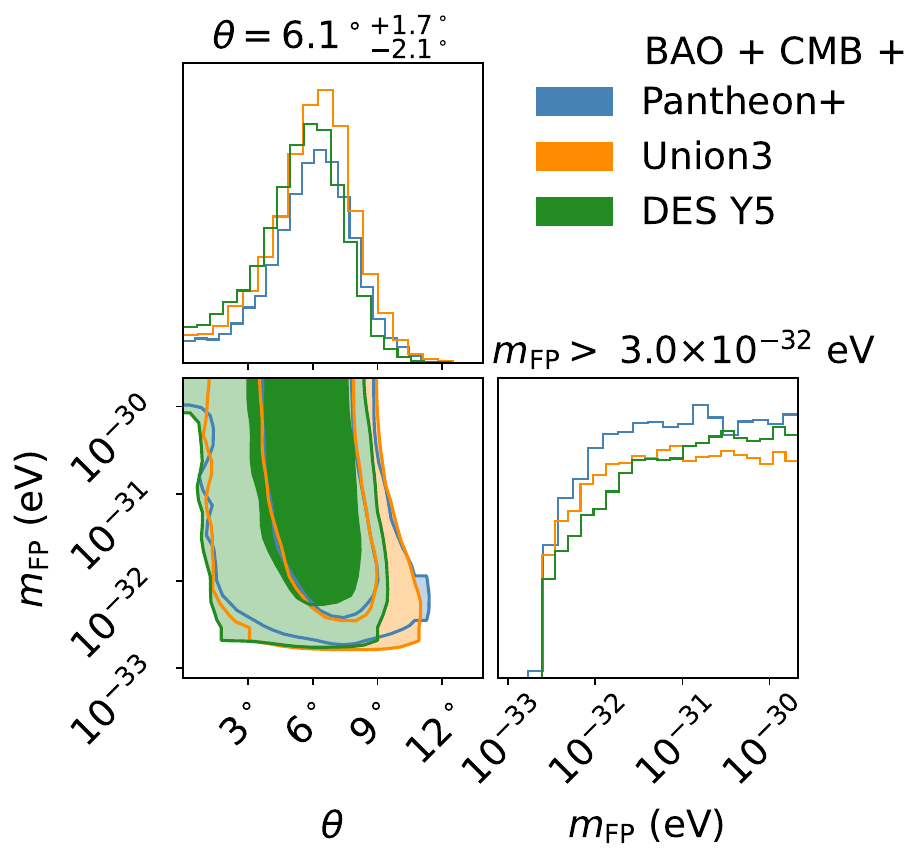}
    \caption{Results for a general bimetric model. Marginalized posterior constraints on $\theta$ and $\mfp$ from BAO (DESI DR2), CMB (\emph{Planck} 2019 + ACT), and three different SNe Ia compilations. Contours show the $68 \, \%$ and $95 \, \%$ credence regions. While the mixing angle $\theta$ is constrained, the mass scale $\mfp$ only exhibits a lower bound $\mfp > 3.0 \times 10^{-32} \, \mathrm{eV}$ at $68 \, \%$ credence. For reference, $H_0 \sim 10^{-33} \, \mathrm{eV}$.}
    \label{fig:theta_mFP}
\end{figure}

Our results can be compared to those of Ref.~\cite{Smirnov:2025yru}, particularly their Fig.~3, revealing some significant differences. For example, our full statistical analysis yields constraints on the mixing angle $\theta$ and mass scale $\mfp$ shown in Fig.~\ref{fig:theta_mFP} which has a very different shape compared with Fig.~3 of Ref.~\cite{Smirnov:2025yru}. 
In particular, in the present work $\mfp$ is only constrained from below, with a lower bound of $\mfp > 3.0 \times 10^{-32} \, \mathrm{eV}$ at $68 \, \%$ credence while Ref.~\cite{Smirnov:2025yru} reports a bounded interval for this parameter. The sharp cutoff in $\mfp$ seen in Fig.~\ref{fig:theta_mFP} is due to the Higuchi bound, ensuring a ghost-free cosmology, see Refs.~\cite{Higuchi:1986py,Fasiello:2013woa,DeFelice:2014nja,Hogas:2021fmr}.
We obtain $\theta = {6.1^\circ}^{+1.7^\circ}_{-2.1^\circ}$, whereas Ref.~\cite{Smirnov:2025yru} reports a somewhat higher value of $\theta \simeq 9^\circ$.
Since Ref.~\cite{Smirnov:2025yru} only fits the equation of state, they report wide constraints on the cosmological constant contribution, $\OmL = 0.70 \pm 0.12$, whereas our full statistical analysis yields much tighter constraints, $\OmL = 0.699 \pm 0.004$, comparable to those in the \cpl DESI analysis \cite{DESI:2025zgx}.\footnote{Recall that here $\OmL$ denotes the asymptotic cosmological constant value that $\Ode$ approaches in the future.}
Finally, Ref.~\cite{Smirnov:2025yru} reports tightly bounded intervals for the two remaining theory parameters (referred to as $A$ and $B$ in that work; $\alpha$ and $\beta$ in Ref.~\cite{Hogas:2021fmr}) whereas we obtain flat posterior distributions for these parameters, indicating that they are essentially unconstrained by the data, with the posterior simply reflecting the assumed prior.\footnote{The parameter $B$ ($\beta$ in Ref.~\cite{Hogas:2021fmr}) has a lower bound imposed by the requirement of a working screening mechanism.}

The difference between our analysis and Ref.~\cite{Smirnov:2025yru} likely reflects the difference in methodologies: by constraining the model through a low-redshift ($z < 2.8$) reconstructed equation of state, the analysis of Ref.~\cite{Smirnov:2025yru} does not explicitly account for the high-redshift behavior of dark energy or guarantee consistency with the CMB. 
The approach of the current paper on the other hand incorporates the full data likelihood across BAO, CMB, and SNe~Ia, providing constraints on the theory parameters which ensures compatibility with observations across all relevant redshifts.
Moreover, the equation-of-state reconstruction on which the results of Ref.~\cite{Smirnov:2025yru} are based is unable to capture the behavior of the general bimetric models which allow for a pole at a specific redshift. See Fig.~\ref{fig:wDE}, and Ref.~\cite{Ozulker:2022slu} for a more general discussion on this phenomenon.

Following Ref.~\cite{DESI:2025zgx}, we quantify the goodness of fit in terms of $\Dchi$, that is, the difference in $\chi^2$ between the best-fit bimetric and \lcdm models.
The significance for the preference of the bimetric model is reported in terms of a number of sigmas, following eq.~(22) of Ref.~\cite{DESI:2025zgx}.
As shown in Tab.~\ref{tab:ResTab} (upper part), bimetric gravity yields a modest improvement in goodness of fit compared to \lcdm, with a preference at the $1 \, \sigma$ level.

\begin{table*}
  \centering
  \renewcommand{\arraystretch}{1.5}
  \begin{tabular}{c@{\hskip 4pt}|@{\hskip 4pt}cl|@{\hskip 4pt}c@{\hskip 4pt}|@{\hskip 4pt}c@{\hskip 4pt}|c|c}
    \hline \hline
    Model & \multicolumn{2}{@{\hskip 4pt}c|@{\hskip 4pt}}{Datasets} & $\theta$ (deg) & $H_0$ (km/s/Mpc) & $\Dchi$ & Significance \\
    \hline
    & & Pantheon+ & ${6.1^\circ}^{+1.7^\circ}_{-2.1^\circ}$ & $69.0 \pm 0.4$ & $-5.0$ & $1.1 \, \sigma$ \\
    Bimetric & BAO + CMB + & Union3 & ${6.1^\circ}^{+1.7^\circ}_{-2.2^\circ}$ & $69.0 \pm 0.4$ & $-4.9$ & $1.0 \, \sigma$ \\
    & & DES Y5 & ${5.6^\circ}^{+1.7^\circ}_{-2.3^\circ}$ & $68.8 \pm 0.4$ & $-4.2$ & $0.9 \, \sigma$ \\
    \hline
    & & Pantheon+ & -- & $68.0 \pm 0.6$ & $-6.4$ & $2.1 \, \sigma$ \\
    $w_0 w_a$CDM & BAO + CMB + & Union3 & -- & $66.2 \pm 0.8$ & $-14.8$ & $3.4 \, \sigma$ \\
    & & DES Y5 & -- & $67.1 \pm 0.6$ & $-16.5$ & $3.7 \, \sigma$ \\
    \hline \hline
    & & Pantheon+(SH0ES) & ${8.0^\circ}^{+1.7^\circ}_{-1.5^\circ}$ & $69.7 \pm 0.4$ & $-11.9$ & $2.4 \, \sigma$ \\
    Bimetric & BAO + CMB + & Union3 + $H_0$ & ${7.7^\circ}^{+1.7^\circ}_{-1.6^\circ}$ & $69.5 \pm 0.4$ & $-10.4$ & $2.1 \, \sigma$ \\
    & & DES Y5 + $H_0$ & ${7.3^\circ}^{+1.5^\circ}_{-1.6^\circ}$ & $69.4 \pm 0.4$ & $-9.2$ & $1.9 \, \sigma$ \\
    \hline
    & & Pantheon+(SH0ES) & -- & $69.2 \pm 0.6$ & $-10.1$ & $2.7 \, \sigma$ \\
    $w_0 w_a$CDM & BAO + CMB + & Union3 + $H_0$ & -- & $69.0 \pm 0.7$ & $-7.7$ & $2.3 \, \sigma$ \\
    & & DES Y5 + $H_0$ & -- & $68.5 \pm 0.5$ & $-10.9$ & $2.9 \, \sigma$ \\
    \hline \hline
  \end{tabular}
  \caption{Results for the general bimetric model, with corresponding results for the \cpl model shown for comparison.
  The upper part of the table lists results without locally calibrated SNe Ia, while the lower part includes local distance-ladder calibration: ``SH0ES'' denotes the inclusion of the 42 locally calibrated SNe~Ia from Pantheon+, and for the Union3 and DES~Y5 combinations, we instead impose a Gaussian prior, $H_0 = 73.0 \pm 1.0 \hu$, from the SH0ES-team calibration, denoted ``$H_0$.''
  Without local calibrators the statistical preference for bimetric gravity over \lcdm is modest, with a significance of roughly $1 \, \sigma$. In contrast, the \cpl model yields a more substantial improvement, leading to a higher significance in the range $2.1$--$3.7 \, \sigma$, depending on the SN Ia compilation. On the other hand, the \cpl model favors lower values of $H_0$, thus exacerbating the Hubble tension, while the bimetric model favors $H_0 = 69.0 \pm 0.4 \hu$, reducing the tension with the local calibration to $3.7 \, \sigma$.
  Including the local SN~Ia calibration, which directly accounts for the impact on the Hubble tension, results in a comparable statistical preference for the bimetric and \cpl models, with a significance of $\simeq 2 \, \sigma$ for the bimetric model and $2$--$3 \, \sigma$ for the \cpl model.
  Our results for the \cpl model differ somewhat from Tab.~VI of Ref.~\cite{DESI:2025zgx}. We attribute this difference to our usage of the \emph{Planck} 2018 + ACT compressed likelihood, whereas Ref.~\cite{DESI:2025zgx} uses the full power of the CMB information, including CMB lensing which increases the significance somewhat. See the discussion in Section~VII and Appendix~A of Ref.~\cite{DESI:2025zgx}. Additionally, our application of the SNe Ia cutoff $z = 0.023$ (instead of $z=0.01$) shifts the models slightly closer to \lcdm results.}
  \label{tab:ResTab}
\end{table*}

\begin{figure*}
    \centering
    \includegraphics[width=1\linewidth]{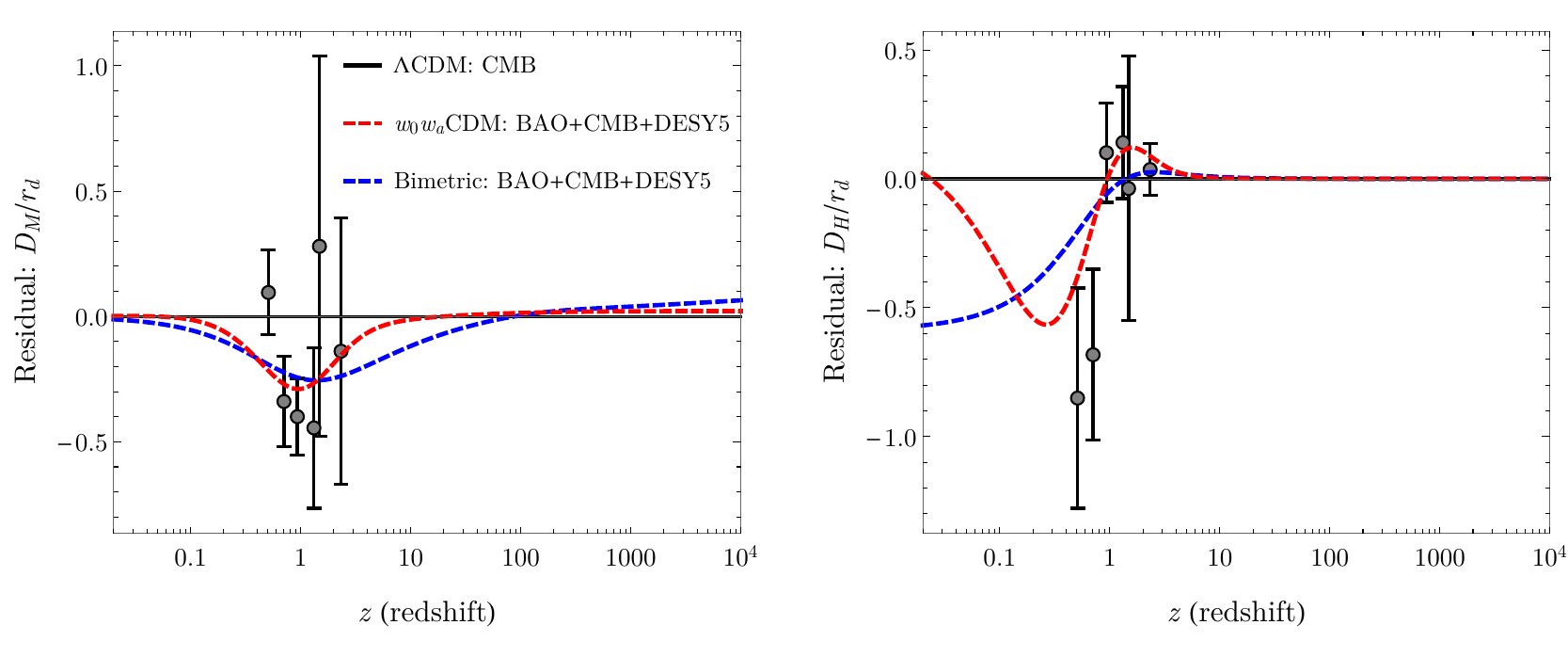}
    \caption{Comparing data to models shown in residual form relative to a fiducial \lcdm model, following the style of Fig.~13 of Ref.~\cite{DESI:2025zgx}. The fiducial \lcdm model is the CMB prediction for the distance ratios. The BAO (DESI DR2) data points are shown with error bars. Here, $D_M$ denotes the angular diameter distance, $D_H$ the Hubble distance, and $r_d$ the sound horizon at the baryon drag epoch. The best-fit \cpl and bimetric models are shown as dashed lines. The primary improvement in the quality of fit of the \cpl model compared with the bimetric model arises from the low-redshift Hubble distance ratios at $z_\mathrm{eff} = 0.51$ and $z_\mathrm{eff} = 0.71$ (right panel). Conversely since the Hubble distance is inversely proportional to the Hubble parameter, $D_H(z) = c / H(z)$, the bimetric model predicts a higher value for $H_0$ compared with the \cpl model, thus reducing the Hubble tension.}
    \label{fig:residuals}
\end{figure*}

\begin{figure}
    \centering
    \includegraphics[width=1\linewidth]{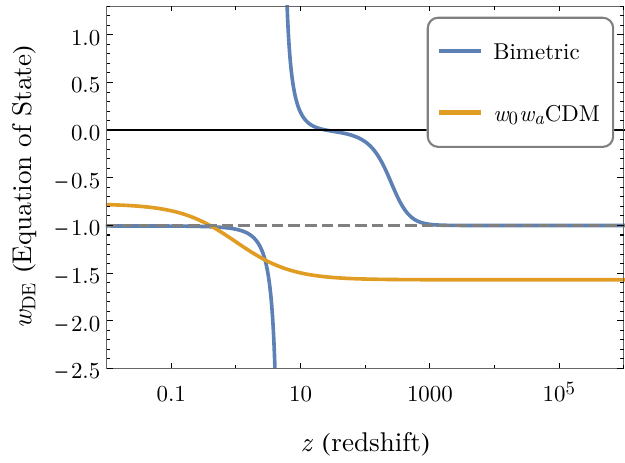}
    \caption{Redshift evolution of the dark energy equation of state for the best-fit bimetric and \cpl models from the combined BAO (DESI DR2), CMB (\emph{Planck} 2018 + ACT), and SNe Ia (DES Y5) data.
    For the \cpl model, $\wde$ is $<-1$ and approximately constant at early times, crosses the phantom divide at $z \simeq 0.4$, and approaches $-0.8$ today.
    For the bimetric model $\wde$, starts at $-1$ in the early universe, then rises monotonically after recombination, with $\Ode$ increasing from a negative value, crossing zero at $z \simeq 5$, where the equation of state formally diverges, and remaining below $-1$ thereafter, steadily approaching $-1$. The divergence in $\wde$ simply reflects the change of sign in the energy density and does not indicate a physical pathology.
    This is a behavior that the CPL parameterization cannot capture. 
}
    \label{fig:wDE}
\end{figure}

In Fig.~\ref{fig:residuals}, we illustrate how the bimetric and \cpl models improve the fit by plotting their residuals relative to a fiducial \lcdm model. This fiducial \lcdm model represents the best fit to the CMB data.
As shown in Fig.~\ref{fig:residuals}, the \cpl model improves the fit by increasing the expansion rate, thus decreasing the Hubble distance, relative to the fiducial \lcdm model at the low-$z$ BAO points ($z_\mathrm{eff} < 0.93$). However, due to the phantom crossing (see Fig.~\ref{fig:wDE}), at redshifts below roughly $0.03$, the expansion rate falls short of the \lcdm prediction from the CMB, thereby worsening the Hubble tension.
In contrast, the bimetric model improves the fit by raising the expansion rate for $z \lesssim 1.4$, with the increased expansion rate continuing all the way to present day ($z=0$). This behavior, illustrated in Fig.~\ref{fig:wDE}, naturally reduces the Hubble tension while providing a better overall fit compared with the \lcdm model.

The inferred Hubble constant from the bimetric model is $H_0 = 69.0 \pm 0.4 \hu$, which is $0.6 \hu$ higher than the value obtain when fitting the same datasets under \lcdm, and $1$--$3 \hu$ higher than that of the \cpl model, depending on the SNe Ia sample.
Notably, the value inferred using the bimetric model is fully consistent with the local distance ladder result from the CCHP team, which uses Tip of the Red Giant Branch (TRGB) stars to calibrate the SNe Ia, yielding $H_0 = 69.8 \pm 0.8 \, (\mathrm{stat}) \pm 1.7 \, (\mathrm{sys}) \hu$ \cite{Freedman:2019jwv}. 
The tension with the SH0ES measurement, based on Cepheid-calibrated SNe Ia, is reduced to $3.7 \, \sigma$, compared to $\simeq 5 \, \sigma$ under the \cpl model---representing a significant alleviation, though not a complete resolution.

In summary, the bimetric model offers a modest improvement, compared to \lcdm, in the fit to DESI DR2 data when combined with CMB and SNe Ia, whereas the \cpl model yields a more significant enhancement in fit quality.
However, unlike the \cpl model—which exacerbates the Hubble tension—the bimetric model simultaneously eases the tension, offering a more coherent picture when considering both fit quality and consistency with local measurements of $H_0$.

To quantify the simultaneous improvement in the Hubble tension and the fit to DESI DR2 + CMB + SNe Ia, we include the 42 locally Cepheid-calibrated SNe Ia from the \pp sample. The results are presented in Tab.~\ref{tab:ResTab} (lower part). For the \uni and \des data samples, we instead impose a Gaussian prior of $H_0 = 73.0 \pm 1.0 \hu$, as per the SH0ES-team measurement \cite{Riess:2021jrx}.
As expected, this reduces the difference in model preference between the bimetric and \cpl models, with the bimetric model being favored over \lcdm at the $2 \, \sigma$ level, while the \cpl model shows a preference at $2$--$3 \, \sigma$ level when the locally calibrated SNe Ia are factored in.

It is worth noting that the BAO distance measurements used here rely on correlations between matter tracers within thick redshift shells. Converting these redshifts into distances requires assuming a cosmological model, universally taken to be \lcdm or a close variant \cite{Carter:2019ulk,Pan:2023zgb,Sanz-Wuhl:2024uvi,DESI:2024ude}. This inevitably introduces a model dependence that, strictly speaking, is incompatible with the alternative expansion histories explored in this work \cite{Sanchez:2010zg,Carvalho:2015ica,Alcaniz:2016ryy,deCarvalho:2017xye,Carvalho:2017tuu,Nunes:2020hzy,deCarvalho:2021azj}. Such a dependence could potentially bias the inferred Hubble constant toward lower values, contributing to the Hubble tension. 
One way to minimize this issue is to use 2D transverse BAO data instead. By counting correlations within very thin redshift shells, the need to assume a background cosmology is greatly reduced, since cosmic evolution can be effectively neglected over such narrow slices \cite{Sanchez:2010zg,Carvalho:2015ica,Alcaniz:2016ryy,deCarvalho:2017xye,Carvalho:2017tuu,Nunes:2020hzy,deCarvalho:2021azj}.
Indeed, several recent studies have shown that replacing standard 3D BAO data with 2D BAO measurements can lead to significantly higher values of $H_0$ for some cosmological models \cite{Li:2019yem,Li:2020ybr,Yang:2021eud,Akarsu:2021fol,Akarsu:2022typ,Akarsu:2023mfb,Gomez-Valent:2024tdb,Hernandez-Almada:2024ost,Dwivedi:2024okk}. This effect is also evident in bimetric gravity: fitting 2D BAO data from SDSS jointly with \emph{Planck} 2018 CMB data and Pantheon+ SNe Ia yields $H_0 = 71.0 \pm 0.9$, consistent with the local SH0ES calibration \cite{Dwivedi:2024okk}.
At present, model-independent 2D BAO data from DESI are not available; however, conducting such data would be of significant interest, as it would allow a direct assessment of how the assumptions underlying the DESI data reduction might impact the inferred Hubble tension.

So far, we have considered the general bimetric model with all five theory parameters. The minimal version of the theory that still ensures a ghost-free cosmological evolution and a functioning screening mechanism involves only three free parameters: $\theta$, $\mfp$, and $\OmL$. This model therefore exhibits the same number of parameters as the \cpl model.
In this minimal scenario, the dark energy density $\Ode$ arises entirely from the interaction between the two metrics, without requiring a cosmological constant in either metric sector.
Notably, for this model a small value of $\Ode$ is technically natural in the sense of 't Hooft, with its value being protected against quantum corrections \cite{tHooft:1979rat}.
A characteristic feature of this model is that its early-universe cosmological constant approaches zero, as opposed to the general model where it can be negative. As a result, the minimal model always exhibits a positive phantom dark energy component, with an equation of state satisfying $\wde \leq -1$ throughout cosmic history.

While the improvement in the fit to the combined BAO (DESI DR2), CMB (\emph{Planck} 2018 + ACT), and SNe~Ia data is slightly smaller for the minimal model compared with the general model, the reduced number of theory parameters leads to a comparable overall preference of about $1 \, \sigma$ over \lcdm.
As in the general case, the minimal bimetric model favors higher values of the Hubble constant than the \cpl model, offering more favorable implications for the Hubble tension compared with \cpl.
Incorporating the locally calibrated SNe~Ia results in a preference for the minimal bimetric model over \lcdm at the $2 \, \sigma$ level.
Thus, also this highly constrained, technically natural model with self-accelerating cosmologies exhibits qualitatively similar behavior to the general bimetric case---most notably, the characteristic transition in $\Ode$ driving late-time acceleration and the preference for a higher Hubble constant.

\section{Discussion}
What constitutes convincing evidence for dynamical dark energy? Here, we use the following criteria:\footnote{These are not meant to be definitive, but they do provide a reasonable basis for our current discussion.}
\begin{enumerate}
    \item \textbf{Empirical support:} Is the dynamical dark energy model significantly favored over \lcdm by multiple independent and well-established datasets (e.g., DESI DR2 and the local measurements of $H_0$)?
    \item \textbf{Theoretical foundation:} Is the model grounded in fundamental physics with a finite-dimensional, physically motivated parameter space, rather than being purely phenomenological?
    \item \textbf{Broader testability:} Does the underlying theory yield additional testable predictions in other observational domains that can distinguish it from standard physics?
\end{enumerate}
For the purposes of this discussion, we take a positive answer to all three questions as indicative evidence of dynamical dark energy.
The \cpl model is moderately favored over \lcdm, at the level of $2$--$4 \, \sigma$, when combining DESI DR2, CMB, and SNe Ia data. However, it worsens the Hubble tension and lacks additional observational support, thus (currently) failing to meet our first criterion. 
By contrast, the bimetric model shows only a modest preference over \lcdm at the $1 \, \sigma$ level when tested against the combined DESI DR2, CMB, and SNe~Ia data. However, it does help easing the Hubble tension, albeit not entirely. Taken together, this yields an overall preference of $\simeq 2 \, \sigma$ compared to \lcdm.
In short, neither model completely satisfies the requirement of \emph{significant} support from \emph{several} independent datasets.

A clear advantage of the bimetric model is its strong theoretical foundation, in contrast to the purely phenomenological nature of the CPL parameterization. Bimetric gravity is unique in the sense that it is the only ghost-free theory that allows a massless and a massive gravitational field to interact consistently. This highly constrained construction permits at most four additional constant parameters, making the theory empirically tractable, allowing these parameters to be directly tested and constrained by observations. 
Thus, the bimetric model clearly satisfies the second of condition.

Closely related to the requirement of a solid theoretical foundation is the third condition: broader testability beyond standard cosmological observations. A purely phenomenological model such as \cpl does not lend itself to independent observational probes---for example, it makes no predictions about gravitational waves or gravitational lensing---because it is agnostic about the underlying physics. By contrast, if dynamical dark energy arises from a fundamental theory, it opens the possibility of testing it through complementary observational channels.

This is precisely the case for bimetric gravity which, for example, predicts distinctive signatures in gravitational waves, including echoes, modified or even inverted waveforms, and beating patterns \cite{Max:2017flc,Max:2017kdc,BeltranJimenez:2019xxx,Brizuela:2023uwt,Brizuela:2025cmz}. While none of these effects have yet been detected, their amplitude and qualitative features depend sensitively on both the mass scale 
$\mfp$ and the mixing angle $\theta$.\footnote{It is worth noting that constraints from the measured propagation speed of gravitational waves only weakly restrict 
$\theta$, since the mixing ensures that there is always a massless mode propagating at the speed of light.} This highlights the potential of dedicated searches---using waveform templates adapted to bimetric gravity---to uncover new physics, potentially even in existing LIGO/Virgo data.

At the same time, the parameter constraints obtained from BAO (DESI DR2) + CMB (\emph{Planck} 2028 + ACT) + SNe Ia remain fully consistent with complementary cosmological observations, as well as tests of gravity on solar-system and galactic scales, thanks to the Vainshtein screening mechanism \cite{Volkov:2011an,Comelli:2011zm,vonStrauss:2011mq,Sjors:2011iv,Volkov:2012wp,Volkov:2012zb,Akrami:2012vf,Volkov:2013roa,Enander:2013kza,Babichev:2013pfa,Koennig:2013fdo,Enander:2015kda,Dhawan:2017leu,Platscher:2018voh,Luben:2018ekw,Hogas:2019ywm,Luben:2020xll,Lindner:2020eez,Caravano:2021aum,Hogas:2021fmr,Hogas:2021lns,Hogas:2021saw,Hogas:2022owf,Guerrini:2023pre,Dwivedi:2024okk}.

To summarize, bimetric gravity presents a theoretically well-founded and broadly testable model that remains fully compatible with existing cosmological and astrophysical constraints, but whose direct empirical support from current data is moderate. It therefore emerges as a strong theoretical candidate, waiting for either more decisive cosmological signals or complementary observational confirmations to establish compelling evidence.

\section{Conclusions}
We have investigated to what extent the latest combination of DESI DR2 BAO measurements, CMB data (\emph{Planck} 2018 + ACT), and SNe~Ia observations (Pantheon+, Union3, and DES Y5) favor dynamical dark energy and whether such a scenario can simultaneously alleviate the long-standing Hubble tension. We have followed two approaches: the widely used phenomenological CPL parameterization (\cpl) and bimetric gravity, a fundamental modification of general relativity that naturally leads to phantom dark energy.

Our results show that while the \cpl model is moderately favored over \lcdm at the $2$--$4 \, \sigma$ level when fitting DESI DR2 + CMB + SNe~Ia, it exacerbates the Hubble tension, leading to even lower inferred values of $H_0$ compared to \lcdm.
By contrast, bimetric gravity yields only a modest improvement in fit quality (around $1 \, \sigma$) relative to \lcdm, but crucially predicts a higher Hubble constant, partially reducing the tension with local distance ladder measurements. Incorporating local calibrator data---either by adding Cepheid-calibrated SNe~Ia directly or imposing the SH0ES prior on $H_0$---brings the overall preference for the bimetric model over \lcdm to the $2 \, \sigma$ level, comparable to that of the \cpl model once the Hubble tension is factored in.

Bimetric gravity is firmly rooted in fundamental theory with a tightly constrained parameter space, and it offers broader testability through distinctive gravitational-wave signatures that could be probed in current or future data. In contrast, purely phenomenological models like \cpl, while flexible in capturing features such as phantom crossing, lack theoretical underpinning and therefore cannot be independently tested outside of cosmological distance measurements.

\begin{acknowledgements}
MH and EM acknowledges support from the Swedish Research Council under Dnr VR 2024-03927. This research utilized the Sunrise HPC facility supported by the Technical Division at the Department of Physics, Stockholm University. The authors used OpenAI’s ChatGPT to assist with editing and proofreading the manuscript.
\end{acknowledgements}

\section*{Data availability}
The distance ratios from DESI DR2, and associated covariance matrix, are publicly available in Tab.~IV of Ref.~\cite{DESI:2025zgx}.
The CMB compressed likelihood from the \emph{Planck} 2018 data release combined with ACT was adopted from Eqs.~(5)--(6) of Ref.~\cite{Bansal:2025ipo}.
\pp data is publicly available at \url{https://github.com/PantheonPlusSH0ES/DataRelease/tree/main/Pantheon%2B_Data}.
\uni data is publicly available at \url{https://github.com/CobayaSampler/sn_data/tree/master/Union3}.
\des data is publicly available at \url{https://github.com/des-science/DES-SN5YR}. 

\appendix
\section{Action and Equations of Motion}
\label{sec:Action_EoM}
In the following we use geometrized units in which the speed of light and the gravitational constant are set to unity meaning that mass, length, and time all have the same units.
The action of bimetric gravity reads \cite{Hassan:2011zd},
\begin{widetext}
\begin{equation}
\label{eq:HRaction}
    \mathcal{S} = \int d^4x \left[ \frac{1}{2\kg} \sqrt{- \det g} \, R_g + \frac{1}{2\kf} \sqrt{- \det f} \, R_f - \sqrt{- \det g} \, \sum_{n=0}^4 \beta_n e_n(S) + \sqrt{- \det g} \, \mathcal{L}_m \right] .
\end{equation}
\end{widetext}
The indices $g$ and $f$ refer to the two metrics with $\kg$ and $\kf$ being the gravitational constants, $R_g$ and $R_f$ the Ricci scalars, $\beta_n$ are the constant theory parameters with dimensions of curvature, $e_n(S)$ are the elementary symmetric polynomials of the square-root matrix $S^\mu{}_\nu = \sqrt{g^{\mu \rho} f_{\rho \nu}}$ \cite{Hassan:2017ugh,higham2008functions}, and $\mathcal{L}_m$ is the matter Lagrangian.
Since \g couples to matter we identify it as the physical metric, determining the space-time geometry of ordinary observers; \f is only observable indirectly via its influence on the physical metric. 

Varying the action \eqref{eq:HRaction} with respect to the two metrics yields the equations of motion:
\begin{subequations}
    \begin{align}
        G_g^\mu{}_{\nu} &= \kg (T^\mu{}_\nu + V_g^\mu{}_\nu) \\
        G_f^\mu{}_{\nu} &= \kf V_f^\mu{}_\nu
    \end{align}
\end{subequations}
where $G_g^\mu{}_{\nu}$ and $G_f^\mu{}_{\nu}$ are the two Einstein tensors, $T^\mu{}_\nu$ is the matter stress--energy and the bimetric stress--energies $V_g^\mu{}_\nu$ and $V_f^\mu{}_\nu$ results from the interaction between the two metrics, and can be expressed as
\begin{subequations}
    \begin{align}
        V_g^\mu{}_{\nu} &= - \sum_{n=0}^3 \beta_n \sum_{k=0}^n (-1)^{n+k} e_k(S)(S^{n-k})^\mu{}_\nu \\
        V_f^\mu{}_{\nu} &= - \sum_{n=0}^3 \beta_{4-n} \sum_{k=0}^n (-1)^{n+k} e_k(S^{-1})(S^{-(n-k)})^\mu{}_\nu .
    \end{align}
\end{subequations}
Assuming the the matter stress--energy is conserved, we get
\begin{equation}
    \nabla_\mu T^\mu{}_\nu = 0
\end{equation}
where $\nabla_\mu$ is the covariant derivative of the physical metric. 
From the Bianchi identity the bimetric conservation law then follows,
\begin{equation}
\label{eq:BRcons}
    \nabla_\mu V^\mu{}_\nu = 0.
\end{equation}

The action \eqref{eq:HRaction} is invariant under the following constant rescaling:
\begin{equation}
\label{eq:rescaling}
    (f_{\mu\nu}, \kf, \beta_n) \to (\omega f_{\mu\nu}, \omega \kf, \omega^{-n/2} \beta_n), \quad \omega = \mathrm{const.}
\end{equation}
meaning that the theory parameters $\beta_n$ are not observables. To circumvent this we introduce the dimensionless rescaling-invariant parameters (remember that we are using geometrized units)
\begin{equation}
    B_n \equiv \kappa_g \beta_n c^n / H_0^2. 
\end{equation}
Here, $c$ is the conformal constant between the two metrics as they approach the final de Sitter phase towards the infinite cosmological future, $f_{\mu \nu}|_{t \to \infty} = c^2 g_{\mu \nu}|_{t \to \infty}$.

Instead of the $B_n$ parameters, one can reparameterize,
\begin{subequations}
    \begin{align}
        \tan^2 \theta &= \frac{B_1 + 3 B_2 + 3 B_3 + B_4}{B_0 + 3 B_1 + 3 B_2 + B_3}, \\
        \mfp^2 &= (B_1 + 2 B_2 + B_3) / \sin^2 \theta, \\
        \OmL &= \frac{B_0}{3} + B_1 + B_2 + \frac{B_3}{3}, \\
        \alpha &= - \frac{B_2 + B_3}{B_1 + 2 B_2 + B_3}, \\
        \beta &= \frac{B_3}{B_1 + 2 B_2 + B_3}.
    \end{align}
\end{subequations}
The advantage of $(\theta, \mfp, \OmL, \alpha, \beta)$, apart from being invariant under the rescaling \eqref{eq:rescaling}, is that they have immediate physical interpretations, as described in the main text.

To get the bimetric Friedmann equation from the equations of motion, we start by assuming homogeneity and isotropy of the two metrics so that the line elements read, in the comoving coordinates of the physical metric,
\begin{subequations}
    \begin{align}
        ds_g^2 &= -dt^2 + a^2_g(t) d\mathbf{x}^2, \\
        ds_f^2 &= -c^2 x^2(t) dt^2 + c^2 a^2_f(t) d\mathbf{x}^2
    \end{align}
\end{subequations}
with $a_g(t)$ and $a_f(t)$ being the scale factors and $x(t)$ the lapse of \f, and $d \mathbf{x}^2 \equiv dx^2 + dy^2 + dz^2$. 
(Spatial curvature can be straightforwardly included, but here we have set it to zero for simplicity.)
With this ansatz the bimetric conservation law \eqref{eq:BRcons} reads
\begin{equation}
    \left( x - a_f / a_g \right) \left[ 1 + 2 \alpha (1-y) + \beta (1-y)^2 \right] = 0
\end{equation}
where an overdot denotes derivative with respect to time, $t$, and $y \equiv a_f / a_g$.
Setting the factor within the square brackets to zero leads to an unhealthy solution \cite{vonStrauss:2011mq,DeFelice:2012mx,Cusin:2015tmf}. Therefore, the expression within the first parenthesis is set to zero.
Doing so we obtain the bimetric Friedmann equation
\begin{equation}
    \left( \frac{H(z)}{H_0} \right)^2 = \Om (1+z)^3 + \Or (1+z)^4 + \Ode(y)
\end{equation}
where
\begin{equation}
    \Ode \equiv \OmL - \sin^2 \theta \, \mfp^2 (1-y) \left[ 1 + \alpha (1-y) + \frac{\beta}{3} (1-y)^2 \right]
\end{equation}
and $y$ is the solution to the following quartic polynomial,
\begin{widetext}
    \begin{align}
	\label{eq:yPoly}
    	& - \frac{1}{3} \cos^2 \theta \, \mfp^2 (1+2\alpha+\beta) \nonumber \\
    	&+ \left[\Omega_\mathrm{tot}(z) + \OmL + \mfp^2 \left(\cos^2 \theta \, (\alpha + \beta) - \sin^2 \theta \, \left(1+\alpha + \frac{\beta}{3}\right)\right)\right] y \nonumber \\ 
    	& + \mfp^2 \left[-\cos^2 \theta \, \beta + \sin^2 \theta \, (1+2\alpha+\beta)\right] y^2 \nonumber\\
    	&-\left[\OmL  + \frac{1}{3} \mfp^2 \left(\cos^2 \theta \, (-1+\alpha-\beta) + 3\sin^2 \theta \, (\alpha+\beta)\right)\right] y^3 \nonumber \\
    	&+ \frac{1}{3} \sin^2 \theta \, \mfp^2 \beta y^4 = 0.
    \end{align}
\end{widetext}
Here, $\Omega_\mathrm{tot}(z) = \Omega_m(z) + \Omega_r(z)$ denotes the total matter density (i.e., pressureless dust plus radiation).

\section{Numerical implementation and likelihood analysis}
\label{sec:NumSol}
\subsection*{Solving the Friedmann equation}
The evolution of the dark energy density $\Ode$ depends the ratio of the scale factors of the two metrics, denoted $y(z)$, so that we can write the modified Friedmann equation:
\begin{equation}
    \label{eq:Friedmann2}
    E^2(z) = \Om (1+z)^3 + \Or (1+z)^4 + \Ode(y(z)),
\end{equation}
where we have defined $E(z) = H(z) / H_0$ and $\Ode(y)$ is a cubic polynomial in $y$, see Ref.~\cite{Hogas:2021fmr} for a detailed expression.
For each redshift, $y(z)$ is found by solving a quartic polynomial equation numerically using Brent's method, implemented via \texttt{scipy.optimize.brentq}. This gives a robust root-finding approach over the physically meaningful interval \(y \in [0,1]\).
In particular, evaluating eq.~\eqref{eq:Friedmann2} at $z=0$ yields a relation that can be used to solve for $\Om$ in terms of the model parameters.

\subsection*{Integrating for distances}
Given the solution for \(E(z)\), we compute cosmological distances by numerically integrating
\begin{equation}
    I(z) = \int_0^z \frac{dz'}{E(z')}
\end{equation}
using a cumulative trapezoidal integration (\texttt{scipy.integrate.cumulative\_trapezoid}) over a logarithmically spaced grid in redshift. The integral is then interpolated with a linear spline (\texttt{scipy.interpolate.interp1d}) for fast evaluations of luminosity-, angular diameter-, and Hubble-distances.

\subsection*{Likelihood}
For the BAO and CMB data points, the log-likelihood is calculated using the relation
\begin{equation}
    \ln \mathcal{L} = -0.5 \left( \chi^2 + \ln \det C + N \ln 2 \pi \right)
\end{equation}
with $C$ being the covariance matrix and $N$ being the number of data points.
The $\chi^2$-value is obtained from
\begin{equation}
    \chi^2 = \mathbf{\Delta}^T C^{-1} \mathbf{\Delta}
\end{equation}
with $\mathbf{\Delta}$ being the residual $\mathbf{\Delta} = \mathbf{y}_\mathrm{data} - \mathbf{y}_\mathrm{model}(\Theta)$, with $\mathbf{y}_\mathrm{data}$ the data vector and $\mathbf{y}_\mathrm{model}(\Theta)$ the model prediction, and $\Theta$ denotes the set of model parameters.

We adopt the following set of model parameters:
\begin{equation}
    \Theta = \left( H_0, \omega_b, \theta, \hat{M}_{\mathrm{FP}}, \omega_\Lambda, \hat{\alpha}, \hat{\beta} \right),
\end{equation}
where $\omega_\Lambda = \Omega_\Lambda h^2$, $\hat{\alpha} = \log_{10}(\alpha + \sqrt{\beta})$, and $\hat{\beta} = \log_{10} \beta$, with $h = H_0 / (100 \, \hu)$. 
The parameters $\alpha$ and $\beta$ are discussed in detail in Ref.~\cite{Hogas:2021fmr}.
A logarithmic scale is applied to some of the parameters to promote efficient exploration of the parameter space over several orders of magnitude.

The mass scale $\hat{M}_{\mathrm{FP}}$ is defined via 
\begin{equation}
    \hat{M}_{\mathrm{FP}} = \log_{10}(\mfp h_{70}),
\end{equation}
where $\mfp$ is the Fierz--Pauli mass in units of $H_0$, and $h_{70} = H_0 / (70 \, \hu)$. 
Equivalently, this can be expressed in physical units as
\begin{equation}
    \hat{M}_{\mathrm{FP}} = \log_{10} \left(\frac{\mfp}{1.49 \times 10^{-33} \, \mathrm{eV}}\right) .
\end{equation}
The total log-likelihood is obtained by summation of the individual datasets, $\ln \mathcal{L}_\mathrm{tot} = \ln \mathcal{L}_\mathrm{BAO} + \ln \mathcal{L}_\mathrm{CMB} + \ln \mathcal{L}_\mathrm{SN}$ and we ensure a ghost-free cosmology and a functioning Vainshtein screening mechanism by checking the corresponding constraints, listed in Ref.~\cite{Hogas:2021fmr}, setting $\ln \mathcal{L}_\mathrm{tot} = -\infty$ whenever the conditions are violated.

The SNe~Ia data vector is the collection of calibrated\footnote{With respect to to stretch, color, host galaxy properties and observational biases.} B-band magnitudes, $m_B$, for each supernova. The theoretical model prediction is given by
\begin{equation}
    m_B(z) = 5 \log_{10} D_L(z) + 25 + M_B
\end{equation}
where $D_L$ is the luminosity distance in units of Mpc and $M_B$ is the fiducial absolute magnitude. 
Here, we marginalize over $M_B$, except when including the local calibrator supernovae from the \pp sample \cite{Goliath:2001af}.

\subsection*{Sampling the posterior}
We sample the posterior with the affine-invariant ensemble sampler implemented in \texttt{emcee}. For HPC runs, the code uses \texttt{schwimmbad} to parallelize via MPI. To ensure convergence, we require that the total chain length exceeds 100 times the estimated autocorrelation time, $\tau$, and perform a visual inspection of the chains as an additional check. The initial $2\tau$ samples are discarded as burn-in and excluded from the analysis. We use twice as many walkers as there are model parameters.

\bibliography{bibliography}{}
\bibliographystyle{apsrev4-1}

\end{document}